\documentclass{sig-alternate-10pt}

\usepackage{url}
\usepackage[export]{adjustbox}
\usepackage{color,soul}
\usepackage{enumitem}
\usepackage{balance}
\usepackage{xspace}

\usepackage{caption}
\usepackage{multirow}
\usepackage{graphicx}
\usepackage{mfirstuc}
\usepackage{subcaption}

\renewcommand{\footnotesize}{\fontsize{6pt}{8pt}\selectfont}

\title{UB-ANC Drone: A Flexible Airborne Networking and Communications Testbed}

\numberofauthors{1}
\author{
\alignauthor Jalil Modares and Nicholas Mastronade\\
	\affaddr{Department of Electrical Engineering}\\
	\affaddr{University at Buffalo}\\
	\email{\{jmod, nmastron\}@buffalo.edu}
}

\begin{document}
\sloppy

\maketitle

\subsection*{Abstract}
We present the University at Buffalo's Airborne Networking and Communications Testbed (UB-ANC). 
UB-ANC is an open software/hardware platform that aims to facilitate rapid testing and repeatable comparative evaluation of airborne networking and communications protocols at different layers of the protocol stack. 
It combines quadcopters capable of autonomous flight with sophisticated command and control capabilities and embedded software-defined radios (SDRs), which enable flexible deployment of novel communications and networking protocols.
This is in contrast to existing airborne network testbeds, which rely on standard inflexible wireless technologies, e.g., Wi-Fi or Zigbee.

UB-ANC is designed with emphasis on modularity and extensibility, and is built around popular open-source projects and standards developed by the research and hobby communities. This makes UB-ANC highly customizable, while also simplifying its adoption. 
In this paper, we describe UB-ANC's hardware and software architecture.

\vspace{-3pt}
\section{Introduction}\label{sec:intro}

Networked unmanned aerial vehicles (UAVs) have emerged as an important technology for public safety, commercial, and military applications including search and rescue, disaster relief, precision agriculture, environmental monitoring, and C3ISR (i.e., command and control, communications, intelligence, surveillance and reconnaissance). 
However, designing, implementing, and testing UAV networks poses numerous interdisciplinary challenges because the underlying communications and networking problems cannot be explored independently of aero-mechanical, sensing, control, embedded systems, and robotics challenges. Indeed, UAV networks are fundamentally {\it cyber-physical systems}~\cite{namuduri2012airborne}.

Although the physical characteristics of a UAV network can be simulated~\cite{le2009reliable, namuduri2012airborne, atl:rohrer,tiwari2008mobility}, actual implementations and field-tests have been recognized as crucial for demonstrating and evaluating solutions in real-world operating environments~\cite{proc:allred, cheng2013characterizing, atl:frew}. 
Unfortunately, there is currently no suitable experimental testbed framework enabling researchers to holistically explore these challenges. To this end, we have developed a software-defined UAV networking platform at the University at Buffalo (UB). The platform, which we call UB's Airborne Networking and Communications Testbed (UB-ANC), combines quadcopters that are capable of autonomous flight with sophisticated command and control capabilities and software-defined radios (SDRs\footnote{Note that UB-ANC's software architecture also accommodates off-the-shelf wireless networking technologies, e.g., Wi-Fi, Zigbee, and LTE.}), which enable flexible deployment of novel communications and networking protocols. In particular, UB-ANC provides us the ability to collect data to measure and understand the connection between the underlying networking and communications capabilities and the ability of the UAVs to effectively accomplish different tasks in different network environments.

In this paper, we describe the design and implementation of UB-ANC. Our contributions are as follows: 
\begin{itemize}
\item We define a modular and extensible open platform with reconfigurable communications and networking capabilities, which can be easily modified for rapidly testing novel protocols at different layers of the protocol stack. UB-ANC not only supports off-the-shelf wireless networking technologies (e.g., Wi-Fi, Zigbee, LTE), but also custom software-defined wireless technologies. To the best of our knowledge, UB-ANC is the first aerial networking testbed that leverages SDR transceivers.
\item We leverage source code from a popular open-source ground station (APM Planner~2) to enable sophisticated command and control capabilities among drones. Unlike conventional setups, where a remote laptop is used as a ground station to monitor and control a drone over a telemetry link, we equip every drone with an on-board embedded computer that runs a simplified version of the ground station software. 
\item The on-board ground station is built around the popular open-source Micro Air Vehicle Communications Protocol (MAVLink), which specifies message formats for communication between ground stations and a MAVLink compatible flight controllers. The on-board ground station can send commands to the on-board flight controller or to other drones in the network. In this way, our platform supports both centralized and distributed mission planning, and allows missions to be planned statically or dynamically.
\item Our proposed framework works with all MAVLink compatible flight controllers. Consequently, UB-ANC not only supports different flight controllers, but also many different types of vehicles including rovers, boats, planes, helicopters, and multirotors.
\item We briefly describe the UB-ANC Emulator, which is an emulation environment for functionally verifying UB-ANC's core software components prior to deployment on the actual drone hardware. 
\end{itemize}

The rest of the paper is organized as follows. 
In Section~\ref{sec:relwork}, we discuss related work. 
In Sections~\ref{sec:hw} and~\ref{sec:sw}, we introduce UB-ANC's hardware and software architectures, respectively. 
In Section~\ref{sec:emul}, we introduce the UB-ANC Emulator.
We conclude the paper in Section~\ref{sec:con}.

\section{Related Work}\label{sec:relwork}

There has been a lot of interest in the research community on UAV networking. In~\cite{atl:rohrer}, Rohrer et al. develop a domain-specific architecture and protocol suite for cross-layer optimization of airborne networks. They introduce a TCP-friendly transport protocol, IP-compatible network layer, and geolocation aware routing. They perform simulations of their protocols in network simulator programs (ns-2 and ns-3). In~\cite{tiwari2008mobility}, the authors propose the Mobility Aware Routing / Mobility Dissemination Protocol (MARP/MDP) to reduce latency and routing overheads by exploiting the known trajectories of airborne nodes. The nodes' trajectories are preplanned to maximize network connectivity using techniques in~\cite{tiwari2008towards}. MARP/MDP is compared against OLSR and AODV using the QualNet network simulator. In~\cite{le2009reliable}, Le et al. simulate a reliable user datagram protocol in OPNET Modeler v14.5 and in~\cite{namuduri2012airborne}, Namuduri et al. discuss cyber-physical aspects of airborne networks and use ns-2 to study average path durations under different node velocities, hop counts, and node densities. While this prior work contributes significantly to the advancement of UAV networking protocols and understanding some cyber-physical aspects of UAV networks, the protocols have not been implemented and tested in a real system.


In~\cite{proc:allred}, Allred et al. study airborne wireless sensor networks for atmospheric, wildlife, and ecological monitoring. They equip airborne nodes with off-the-shelf 802.15.4-compliant Zigbee radios. They perform experiments to evaluate the performance of air-to-air, air-to-ground, and ground-to-ground wireless links, as well as network connectivity. In~\cite{atl:frew}, researchers at the University of Colorado test the performance of off-the-shelf IEEE 802.11b (Wi-Fi) networking equipment in an airborne mesh network. They show that a mesh network can extend the communication range among airborne nodes in a small unmanned aerial system (UAS), they explore how a mesh network can be used to enable a remote operator to send command and control information to distant aircraft and to receive telemetry information back over the network, and they use controlled mobility to enable ferrying of delay-tolerant data between nodes in a fractured/partitioned network.

Recently, researchers have started investigating the benefits of equipping drones with SDRs~\cite{dos2014small, jakubiak2015cellular, zhou2014future}. In~\cite{dos2014small}, dos Santos et al. design and implement a drone equipped with a low-cost SDR receiver, which can automatically track wildlife tagged with very high frequency (VHF) radio collars. In~\cite{jakubiak2015cellular}, Jakubiak implements a drone equipped with a low-cost SDR receiver to gather data about the coverage of a cellular network. In~\cite{zhou2014future}, Zhou envisions a system for railways in which drones relay data for passengers in high-speed trains to different networks (e.g., satellite or cellular). While~\cite{dos2014small, jakubiak2015cellular} implement systems with only SDR receivers, ~\cite{zhou2014future} does not provide any system implementation.

In summary, while a lot of significant contributions have been made in designing and implementing UAV networks, which exploit communications and networking technologies for command and control, telemetry, and coordination among multiple agents, existing system implementations rely on inflexible off-the-shelf transceivers or SDR receivers. In contrast, UB-ANC provides a flexible and highly reconfigurable airborne networking and communications platform for designing, implementing, and testing state-of-the-art communications and networking protocols in conjunction with sophisticated mission planning algorithms. While the proposed framework is designed to be compatible with off-the-shelf wireless interfaces, e.g., Wi-Fi, Zigbee, and LTE, to the best of our knowledge, UB-ANC is the first UAV networking platform designed to support SDR transceivers. In general, this provides researchers more flexibility to design, implement, and test new communications and networking protocols for UAVs.
\vspace{-2pt}
\section{Hardware Components}
\label{sec:hw}


In this section, we describe the high-level hardware architecture of a UB-ANC drone. We introduce the core components of a drone that are required to use the UB-ANC platform, while also showing that UB-ANC is flexible and can work in numerous configurations.

There are three main hardware components on-board a UB-ANC drone: a flight controller, an embedded computer, and a wireless network element. We currently use two unique drone configurations as outlined in Table~\ref{tab:configs}, although many others are possible. Both configurations use a Pixhawk{\small \footnote{\url{http://copter.ardupilot.com/wiki/common-pixhawk-overview/}}} flight controller; however, as we will see in Section~\ref{sec:sw}, UB-ANC's software architecture is compatible with many other popular flight controllers. Note that, in both configurations, the Pixhawk is connected to the embedded computer through a USB interface.

The differences between our two drone configurations arise from the choice of wireless network technology. The first configuration uses a USRP E310 SDR\footnote{\url{https://www.ettus.com/product/details/E310-KIT}} from Ettus Research for communication; however, other embedded SDRs could be used instead (e.g., the USRP B200-mini\footnote{\url{https://www.ettus.com/product/details/USRP-B200mini-i}} or the bladeRF\footnote{\url{http://www.nuand.com/}}). 
The USRP E310 includes a 667 MHz dual-core ARM Cortex-A9 processor; therefore, the USRP E310 also servers as the embedded computer. This configuration is designed for developing new communications and networking protocols for UAVs. 

The second configuration uses a Wi-Fi module for communication and a Raspberry Pi 2 as the embedded computer; however, other wireless network technologies (e.g., Zigbee or LTE) and other embedded computers (e.g., Beagleboard\footnote{\url{https://beagleboard.org/}} or ODROID\footnote{\url{http://magazine.odroid.com/odroid-xu4/}}) could be used instead. This configuration is best suited for applied UAV networking research where the focus is not on the specific communications and networking protocols, but on using multiple networked UAVs to accomplish a task.

\begin{figure}[t]
\begin{center}
\includegraphics[max width=\linewidth]{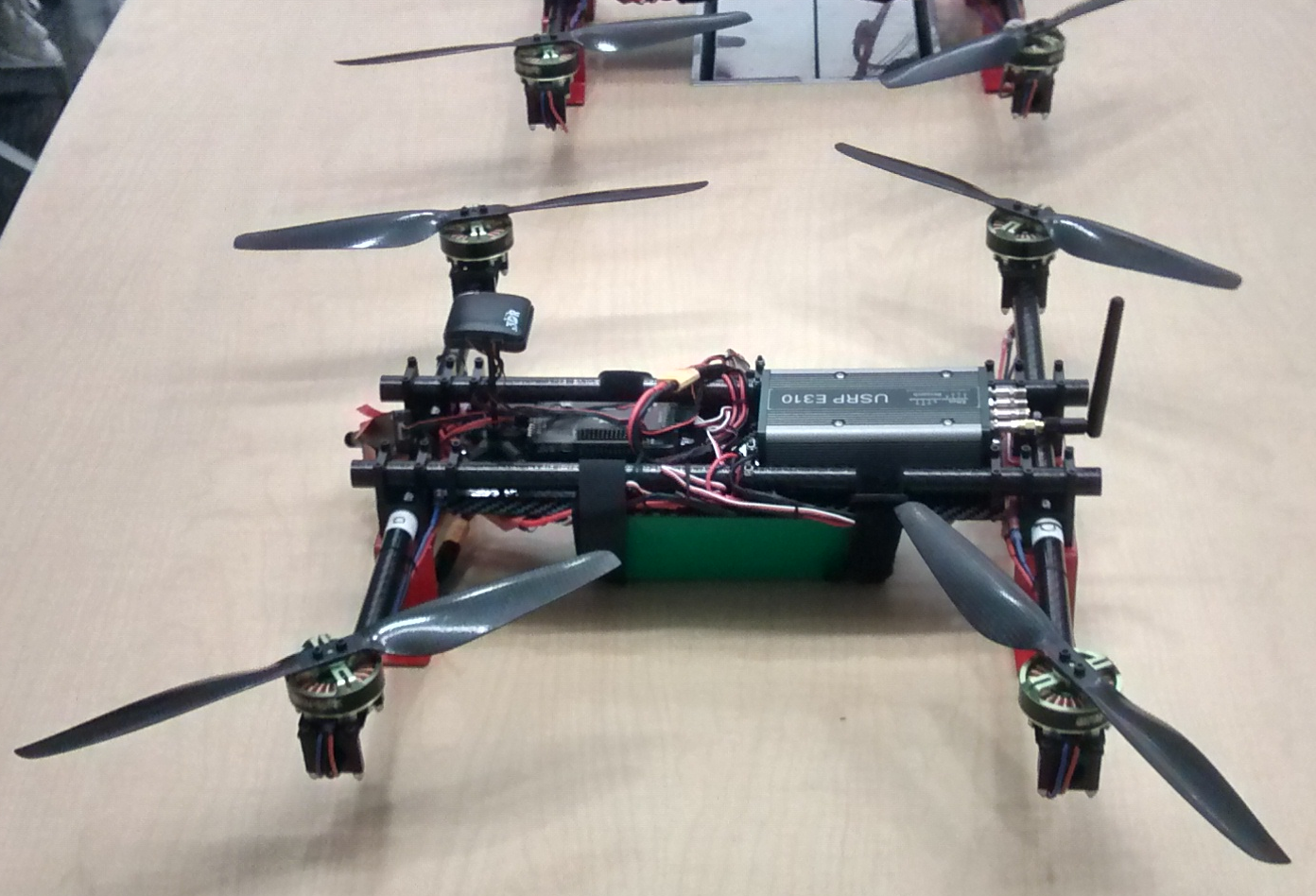}
\end{center}
\vspace{-7pt}
\caption{A UB-ANC drone (SDR configuration).}
\label{fig:ub-anc}
\end{figure}

Figure~\ref{fig:ub-anc} shows one of our three custom-built UB-ANC drones in the SDR configuration. It achieves over 25 minutes of flight time while carrying the 400 g USRP E310 as its payload (for a total weight of 3.125 kgs).

\begin{table*}[t]
\caption{Comparison between two UB-ANC drone configurations.}
\vspace{-7pt}
\begin{center}
\label{tab:configs}
\resizebox{\textwidth}{!}{%
\begin{tabular}{ | l | c | c | }
\hline
                             & \textbf{SDR Configuration}          & \textbf{Wi-Fi Configuration}             \\
\hline
\textbf{Flight Controller}   & Pixhawk                             & Pixhawk                                  \\
\hline
\textbf{Embedded Computer}   & USRP E310 / Dual Core ARM Cortex-A9 & Raspberry Pi 2 / Quad-Core ARM Cortex-A7 \\
\hline
\textbf{Wireless Technology} & USRP E310 SDR                       & Wi-Fi                            \\
\hline
\end{tabular}%
}
\end{center}
\end{table*}

\vspace{-2pt}
\section{Software Components}\label{sec:sw}

Now that we know the hardware requirements of a UB-ANC drone, we are ready to describe UB-ANC's software architecture. Recall from Section~\ref{sec:hw} that a UB-ANC drone includes a flight controller and an embedded computer. In our setup, the embedded computer runs Yocto Linux\footnote{\url{https://www.yoctoproject.org}} as its operating system and the flight controller runs ArduPilot APM:Copter\footnote{\url{http://copter.ardupilot.com}} as its firmware.
The systems are connected to each other using USB CDC-ACM as a serial port with baud rate 115200~bps.

\begin{figure*}[t]
	\begin{center}
	\includegraphics[max width=\linewidth]{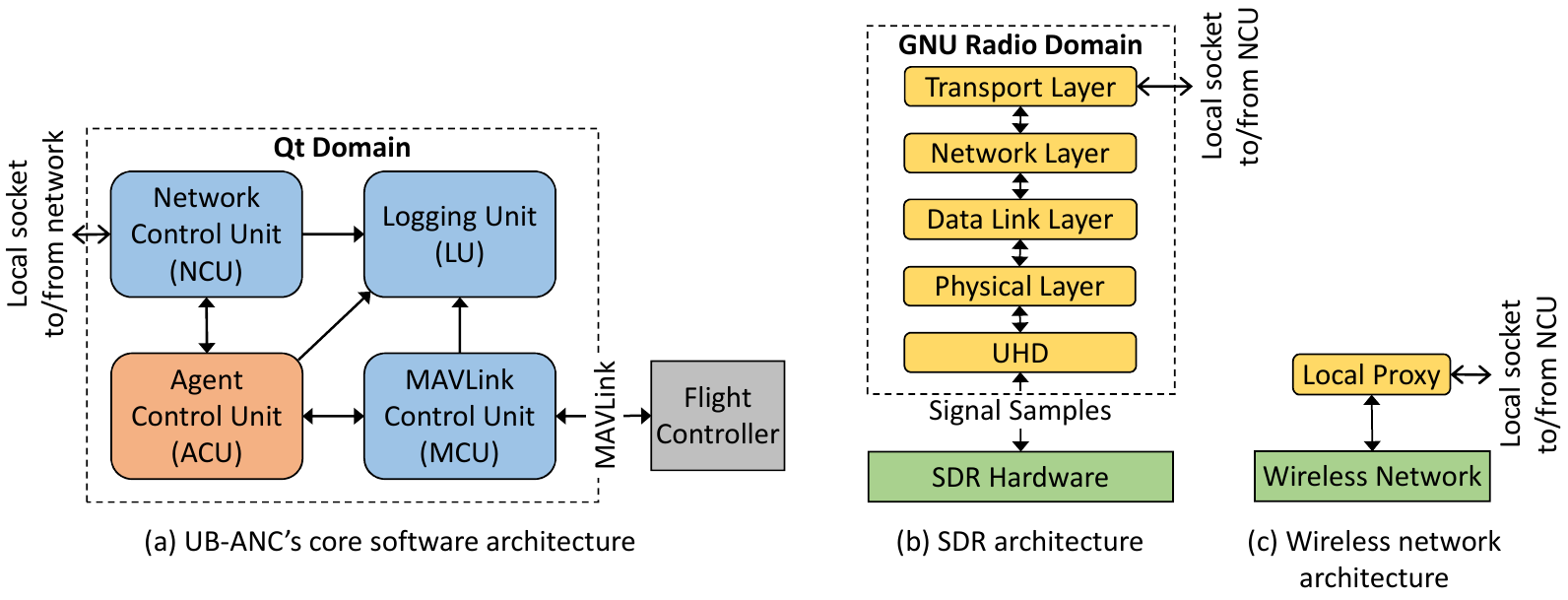}
	\end{center}
    \vspace{-7pt}
	\caption{High-level software architecture diagram. (a) UB-ANC's core software architecture with its interface to the network. (b) SDR architecture with its interface to the Network Control Unit. (c) Standard wireless network architecture with its interface to the Network Control Unit.}
	\label{fig:arch}
\end{figure*}

Figure~\ref{fig:arch}(a) provides a high-level diagram of UB-ANC's core software architecture, which comprises four components: the Agent Control Unit (ACU), the Network Control Unit (NCU), the MAVLink Control Unit (MCU), and the Logging Unit (LU). The ACU is the ``brains'' of a UB-ANC drone: it contains the mission planning logic and interfaces with (i) the NCU to talk with different network elements; (ii) the MCU to talk with different flight controllers; and (iii) the LU to log status information. Table~\ref{tab:API} provides details about the APIs that the ACU uses to interface with the NCU and the MCU. Note that the list of methods in Table~\ref{tab:API} is illustrative, but not exhaustive.

The aforementioned software components are implemented using Qt\footnote{\url{http://www.qt.io}}, which is an object-oriented C++ cross-platform application development framework. We have chosen Qt as the main application framework based on the following considerations:
\begin{itemize}
\item It facilitates event-driven programming and makes it easy to maintain a modular design. Specifically, using Qt's signals and slots mechanism, components can communicate by emitting signals and capturing other components' signals using slots.
\item It is a stable open-source application framework that has been used in many other open-source projects. In particular, some of the open-source software that we are reusing in this project is already implemented using Qt.
\item It is a C++ object-oriented framework, which facilitates efficient coding while maintaining high-performance operation.
\item It is cross-platform, which makes it easy to port the project across different operating systems, like Windows CE, Custom Embedded Linux, Android, and iOS.
\end{itemize}

\begin{table*}[t]
\centering
\caption{Abbreviated front-end APIs for the Network and MAVLink Control Units (i.e., the NCU and MCU).}
\vspace{-5pt}
\resizebox{\textwidth}{!}{%
\begin{tabular}{|c|c|l|l|}
\hline
\textbf{Component}                     & \textbf{Class}                 & \multicolumn{1}{c|}{\textbf{Method}} & \multicolumn{1}{c|}{\textbf{Description}}                              \\ \hline
\multirow{7}{*}{Network Control Unit}          & \multirow{3}{*}{UBNetwork}     & getData()                            & Return data from the receive buffer                                    \\ \cline{3-4} 
                                       &                                & sendData()                           & Send data to the send buffer                                           \\ \cline{3-4} 
                                       &                                & dataReady()                          & A Qt signal emitted when data is in the receive buffer         \\ \cline{2-4} 
                                       & \multirow{4}{*}{UBPacket}      & setSrcID()/getSrcID()                & Set/get the source MAV ID for the packet                               \\ \cline{3-4} 
                                       &                                & setDesID()/getDesID()                & Set/get the destination MAV ID for the packet                          \\ \cline{3-4} 
                                       &                                & setPayload()/getPayload()                & Set/get the payload for the packet                          \\ \cline{3-4} 
                                       &                                & packetize()/depacketize()            & Make/parse the packet stream                                     \\ \hline
\multirow{11}{*}{MAVLink Control Unit} & UASManager                     & UASCreated()                         & A Qt signal emitted when a new flight controller is detected   \\ \cline{2-4} 
                                       & \multirow{3}{*}{LinkManager}   & getLink()                            & Return the ID of the specific link                                     \\ \cline{3-4} 
                                       &                                & getLinkType()                        & Return the type of the link (Serial, TCP, ...)                         \\ \cline{3-4} 
                                       &                                & connectLink()                        & Connect to the specific link                                           \\ \cline{2-4} 
                                       & \multirow{5}{*}{UASInterface}  & setMode()                            & Set the mode of the flight controller                                  \\ \cline{3-4} 
                                       &                                & getAltitude()                        & Return the quad-rotor's altitude                                       \\ \cline{3-4} 
                                       &                                & setHeartbeatEnabled()                & Enable HEARTBEAT message to the flight controller          \\ \cline{3-4} 
                                       &                                & executeCommand()                     & Send a specific MAVLink command to the flight controller               \\ \cline{3-4} 
                                       &                                & isArmed()                   		   & Returns 1 if the flight controller is armed; 0 otherwise. \\ \cline{2-4} 
                                       & \multirow{2}{*}{LinkInterface} & setPortName()                        & Specify the serial port                                                \\ \cline{3-4} 
                                       &                                & setBaudRate()                        & Set the baud rate of the serial port                                   \\ \hline
\end{tabular}%
}
\label{tab:API}
\end{table*}

Before we describe each software component in detail, we highlight the key features of the software architecture design:
\begin{itemize}
\item {\bf Modularity:} UB-ANC's software architecture is designed to be modular. Each component has a well-defined task so that it can be easily modified and debugged. 
\item \textbf{Extensibility:} The components have well-defined interfaces allowing for easy extensibility. For instance, the NCU and MCU have well-defined front-end and back-end interfaces that allow them to work with different network technologies and different flight controllers, respectively.
\item \textbf{Utilizing popular open-source standards:} As noted in the introduction, UB-ANC leverages the popular MAVLink protocol; therefore,it supports all MAVLink compatible vehicle controllers including APM{\small \footnote{\url{http://ardupilot.org/copter/docs/common-apm25-and-26-overview.html}}}, Pixhawk{\small \footnote{\url{http://copter.ardupilot.com/wiki/common-pixhawk-overview/}}}, Emlid's NAVIO{\small \footnote{\url{http://copter.ardupilot.com/wiki/common-navio-overview/}}}, and Qualcomm's Snapdragon{\small \footnote{\url{http://copter.ardupilot.com/wiki/common-qualcomm-snapdragon-flight-kit/}}}. Moreover, since many vehicle controllers that are designed for rovers, boats, planes, helicopters, and multirotors are based on MAVLink, the UB-ANC platform can be easily deployed on different types of vehicles.
\end{itemize}
In the following subsections, we describe each software component in detail.

\subsection{Agent Control Unit (ACU)}
\label{sec:agent}
The ACU is responsible for any mission that the drone is supposed to complete. It includes the internal logic for deciding what commands to send to the flight controller (through the MCU) and what information to send to other nodes (through the NCU) to accomplish its mission. In general, the mission planning logic can make decisions based on local state information and information received from other nodes.

The following code shows a finite state machine algorithm for a simple mission where a drone takes off, loiters (i.e., hovers in position), sends a message to another drone, and then lands. The ACU continuously checks the state of the drone and the mission through a function called {\tt missionTracker}, which is called every 10 milliseconds (100~Hz). Each time the {\tt missionTracker} function is called, the ACU checks if the flight controller is armed and then it executes the appropriate function based on the current state of the mission, i.e., {\tt stageStart()}, {\tt stageLoiter()}, or {\tt stageStop()}.
A portion of the {\tt stageLoiter()} method is also given below, where {\tt executeCommand()} is used to tell the flight controller to land after the loiter time exceeds a threshold. When the drone finishes loitering, it sends a message to a another drone instructing it to start its own simple mission (i.e., takeoff, loiter, and land). 

{\tt \small
\begin{verbatim}
void UBAgent::missionTracker() {
    if (!m_uav->isArmed()) {
        return;
    }
    switch (m_stage) {
    case STAGE_START:
        stageStart();
        break;
    case STAGE_LOITER:
        stageLoiter();
        break;
    case STAGE_STOP:
        stageStop();
        break;
    }
}

void UBAgent::stageLoiter() {
  if ((QGC::groundTimeSeconds() -
         m_loiter_timer > LOITER_TIME)) {
    m_uav->executeCommand(MAV_CMD_NAV_LAND,
      1, 0, 0, 0, 0, 0, 0, 0, 0);
    m_net->sendData(&m_msg);
    m_stage = STAGE_STOP;
    return;
  }
  ...
}
\end{verbatim}
}

\subsection{Network Control Unit (NCU)}
\label{sec:net}
%
%
%
As mentioned earlier, the ACU uses the NCU to send/receive data over the network. For example, one drone can send commands to another drone to visit specific GPS waypoints or, for more sophisticated applications, drones can exchange local state information that their ACUs can use for centralized or distributed mission planning. The NCU is designed so that the underlying network technology can be easily changed while keeping the rest of the system the same. Therefore, we can easily test different wireless network technologies with the same ACU logic so that we can fairly compare the system performance across different configurations.

The NCU provides a front-end API that the ACU uses to access the network. This API comprises the {\tt UBNetwork} and {\tt UBPacket} classes as shown in Table~\ref{tab:API}. The NCU's back-end uses an interprocess communication (IPC) mechanism (a local socket) with a well-defined packet format (Source MAV ID, Destination MAV ID, Payload) to connect to the wireless network. Thus, the NCU can be viewed as the application layer in the network protocol stack. 
%
%
%

The NCU's back-end interface is shown in Figure~\ref{fig:arch}(a) as a bi-directional arrow labeled ``Local socket to/from network.'' While the NCU and its front/back-end interfaces are well-defined, everything beyond the back-end depends on the underlying network technology (e.g., Wi-Fi, Zigbee, LTE, or a software-defined technology). For example, in Figure~\ref{fig:arch}(b), we show how the NCU interfaces with an SDR where the transport, network, data link/MAC and physical layers are implemented within GNU Radio~\cite{malsbury2013modular}. As another example, in Figure~\ref{fig:arch}(c), we show how the NCU interfaces with a local proxy, which uses the existing networking infrastructure of the operating system to connect to a standard wireless network (e.g., Wi-Fi, Zigbee, or LTE). In both Figures~\ref{fig:arch}(b) and \ref{fig:arch}(c), the connection to the NCU is shown as a bi-directional arrow labeled ``Local socket to/from NCU.'' Note that, while the back-end of the NCU that connects to the local proxy is well-defined, the interface from the local proxy to the wireless network is specific to the underlying wireless network technology.

The ACU uses the NCU to send/receive data over the network as follows. When the ACU sends a packet to the NCU, the NCU puts the packet into a private queue called {\tt m\_send\_buffer} and then sends the packet to the wireless network using the aforementioned IPC mechanism.
When a packet is received by the NCU from the network, it raises a signal ({\tt dataReady()}) to notify the ACU that there is a packet in the {\tt m\_receive\_buffer} buffer. The ACU then reads the buffer and processes the received packet. The following code shows how the {\tt sendData()} and {\tt getData()} methods are implemented in the {\tt UBNetwork} class using a Qt container to buffer and unbuffer the data. Notice that, before a packet is queued at the sender, it is first packetized using methods from the {\tt UBPacket()} class.


{\tt \small
\begin{verbatim}
void UBNetwork::sendData(quint8 desID,
       const QByteArray& data) {
    UBPacket packet;
    packet.setSrcID(m_id);
    packet.setDesID(desID);
    packet.setPayload(data);
    QByteArray* stream = 
      new QByteArray(packet.packetize());
    m_send_buffer.enqueue(stream);
    ...
}

QByteArray UBNetwork::getData() {
    QByteArray data;
    if (m_receive_buffer.isEmpty())
        return data;
    QByteArray* stream = m_receive_buffer.dequeue();
    data = *stream;
    delete stream;
    return data;
}
\end{verbatim}
}



\subsection{MAVLink Control Unit (MCU)}
\label{sec:MCU}
The MCU provides a front-end API that the ACU uses to send commands to (and receive messages from) a flight controller. The back-end of the MCU supports different types of connections to the flight controller (e.g., USB, Ethernet, and serial) and can even connect to multiple flight controllers simultaneously\footnote{In general, it is possible for a vehicle to have multiple controllers. For example, a vehicle that can switch between air, land, and water may have a separate controller for each modality.}. The MCU communicates with the flight controller using the MAVLink messaging protocol\footnote{\url{http://qgroundcontrol.org/mavlink/start}};
consequently, the MCU can easily interface with any MAVLink compatible flight controller. 



MAVLink supports various messages and commands\footnote{\url{https://pixhawk.ethz.ch/mavlink}}. One of the most important messages, called the HEARTBEAT, is generated by the MCU and flight controller every second (1~Hz). The HEARTBEAT message shows that the link between the MCU and flight controller is still alive. If the HEARTBEAT message from the MCU is lost, then the flight controller goes into a preconfigured failsafe mode (either return-to-launch, which requires a GPS lock, or land, which does not). On the other hand, the HEARTBEAT message from the flight controller contains information that the MCU can use for different tasks. This information includes, but is not limited to, the type of micro air vehicle (quadcopter, helicopter, fixed wing, etc.); the type of flight controller (APM, Pixhawk, etc.); the mode of the flight controller (armed, autonomous, manual, stabilize, etc.); and the MAVLink protocol version.
Note that not all MAVLink commands are supported by all flight controllers. Therefore, knowledge of the specific type of flight controller is important to ensure that only the correct commands are used.

Table~\ref{tab:mav:com} shows an abbreviated list of some important MAVLink commands. Every MAVLink command is associated with up to seven parameters. For illustration, the parameters of the loiter command ({\tt MAV\_CMD\_NAV\_LOITER\_TIME}), which include the loiter duration, latitude, longitude, and altitude, are shown in Table~\ref{tab:mav:com:param}. The ACU uses the {\tt executeCommand()} method to send specific MAVLink commands to the flight controller (see Table~\ref{tab:API}). A code snippet in Section~\ref{sec:agent} shows how to use the {\tt executeCommand()} method.

\begin{table*}[t]
\caption{An abbreviated list of MAVLink commands.}
\vspace{-8pt}
\begin{center}
\resizebox{\textwidth}{!}{%
\begin{tabular}{ | l | l | l | }
\hline
\textbf{CMD ID} & \textbf{Command Name} & \textbf{Description} \\ \hline
16 & MAV\_CMD\_NAV\_WAYPOINT & Navigate to a waypoint \\
19 & MAV\_CMD\_NAV\_LOITER\_TIME & Loiter around a waypoint for X seconds \\
20 & MAV\_CMD\_NAV\_RETURN\_TO\_LAUNCH & Return to launch location \\
21 & MAV\_CMD\_NAV\_LAND & Land at location \\
22 & MAV\_CMD\_NAV\_TAKEOFF & Takeoff from ground \\
176 & MAV\_CMD\_DO\_SET\_MODE & Set system mode \\
183 & MAV\_CMD\_DO\_SET\_SERVO & Set a servo to a desired PWM value \\
\hline
\end{tabular}
}
\end{center}
\label{tab:mav:com}
\end{table*}

\begin{table}[t]
\centering
\caption{Parameters for the loiter command.}
\label{tab:mav:com:param}
\vspace{-5pt}
\resizebox{\columnwidth}{!}{%
\begin{tabular}{ | l | l | }
\hline
\textbf{Param No.} & \textbf{Description} \\ \hline
1 & Seconds (decimal) \\
2 & Empty \\
3 & Radius around the waypoint, in meters \\
4 & Desired yaw angle \\
5 & Latitude \\
6 & Longitude \\
7 & Altitude \\
\hline
\end{tabular}
}
\end{table}

The MCU is implemented using four classes from an open-source project
%
called APM Planner~2, namely, {\tt UASManager}, {\tt LinkManager}, {\tt UASInterface}, and {\tt LinkInterface}.\footnote{\url{https://github.com/diydrones/apm_planner}} APM Planner~2 is a GUI-based ground station that can be used to define missions, send missions to a flight controller, and track a drone on a map. It is based on Qt and works with MAVLink compatible flight controllers.
As we noted in the introduction, GUI-based ground stations like APM planner~2 are typically loaded on a laptop to monitor and control a drone over a telemetry link; however, in order to support more sophisticated mission planning algorithms than conventional setups (which rely on centralized control), we load ground station software directly onto each drone's embedded computer (enabling fully distributed control).
To achieve this, we carefully stripped away the GUI-based elements of the aforementioned classes to create a light-weight console-based ground station. 

\textbf {LinkManager and LinkInterface:}
The {\tt LinkManager} class is responsible for managing different kinds of links between the flight controller and the MCU (serial link, TCP/UDP link, telemetry link, etc.). Every link has a corresponding class ({\tt SerialLinkInterface}, {\tt TCPLink}, {\tt UDPLink}, etc., which are all derived from the base link class {\tt LinkInterface}). When a link is established, the {\tt LinkManager} creates the corresponding link object. The ACU then uses the link object to control the link (connect, disconnect, set baud rate, etc.). 


\textbf {UASManager and UASInterface:}
The {\tt UASManager} class is responsible for managing different kinds of flight controllers (APM, Pixhawk, etc.). When the MCU receives a HEARTBEAT message, the {\tt UASManager} first determines the type (and ID) of the flight controller that sent the message. If the corresponding flight controller's object does not already exist, then the {\tt UASManager} creates the appropriate flight controller object ({\tt ArduPilotMegaMAV}, {\tt PxQuadMAV}, etc., which are all derived from the base flight controller class {\tt UASInterface}) and puts it in a private list called {\tt m\_uas\_list}. The ACU then uses the flight controller object to send commands to (and receive messages from) the corresponding flight controller.

\subsection{Logging Unit (LU)}
\label{sec:LU}
There is a lot of information that can be tracked in the system including, but not limited to, GPS position (longitude, latitude, and altitude), MAVLink messages, drone ground speed, packet information (e.g., packet ID, source ID, and destination ID), channel state information, etc. We track data in our system using QsLog\footnote{\url{https://github.com/victronenergy/QsLog}}, which is a system logger based on Qt's {\tt QDebug} class. The data can be logged on a MicroSD card so it can be analyzed offline, or it can be sent to a ground station where it can be viewed and analyzed in real-time. The Logging Unit can be configured to provide different levels of verbosity using different logging functions, e.g., {\tt QLOG\_ERROR()}, {\tt QLOG\_WARN()},  and {\tt QLOG\_DEBUG()}.

In Table~\ref{tab:log}, we show an abbreviate log for the simple takeoff, loiter, and landing mission described in Section~\ref{sec:agent}, which we tested on one of our UB-ANC drones. The log shows time stamps for key events (with millisecond granularity) along with the corresponding event descriptions. In Table~\ref{tab:log}, we see that the flight controller is initialized to the ``Stabilize'' mode (a simple manual flight mode) and then its barometer is calibrated. After some delay, the motors are manually armed using an RC remote, which triggers the autonomous mission to start. Once armed, the quadcopter takes off and climbs in altitude. After approximately 9 seconds, it switches to ``Loiter'' mode and hovers for approximately 25 seconds before it switches to ``Land'' mode. The mission ends when the drone lands. 

\begin{table}[tbp]
\centering
\caption{Abbreviated mission log.}
\vspace{-5pt}
\label{tab:log}
\resizebox{\columnwidth}{!}{%
\begin{tabular}{|l|l|}
\hline
\multicolumn{1}{|c|}{\textbf{Time Stamp}} & \multicolumn{1}{c|}{\textbf{Event Description}} \\ \hline
2016-02-08T16:19:57.637                   & Mode changed to Stabilize                       \\ \hline
2016-02-08T16:19:57.639                   & Calibrating barometer                           \\ \hline
2016-02-08T16:20:46.188                   & Arming motors                                   \\ \hline
2016-02-08T16:20:54.813                   & Mode changed to Loiter                          \\ \hline
2016-02-08T16:21:19.406                   & Mode changed to Land                            \\ \hline
2016-02-08T16:21:31.773                   & Mission complete                                \\ \hline
\end{tabular}
}
\end{table}

\vspace{-2pt}
\section{UB-ANC Emulator}
\label{sec:emul}

Experimentation with UAV networks is very challenging. This is not only because of the complexity of the involved systems, but also because experimentation requires multiple operational drones, suitable weather conditions, well-trained personnel, many charged batteries, and adherence to continuously evolving regulations. For these reasons, it is essential that time spent in the field is not wasted because of errors in the software design and implementation.

To address this challenge, we have developed the UB-ANC Emulator, which is an open-source emulation environment for verifying the functionality of UB-ANC's core software components (i.e., the ACU, NCU, MCU, and LU illustrated in Figure~\ref{fig:arch}(a)) in software prior to deployment. UB-ANC Emulator is designed to facilitate rapid transition from emulation to experimentation. In particular, it uses (i) the same software and interfaces that run on the actual drone hardware (including the MAVLink protocol), (ii) a software-in-the-loop (SITL) simulator of the flight controller's firmware, and (iii) a GUI for visualizing emulated missions with multiple networked UAVs. Thus, if the core software components function correctly in the emulator, then they will also work when deployed on the drone hardware.
\vspace{-2pt}
\section{Conclusion and Future Work}\label{sec:con}

We introduced the hardware and software architecture of the University at Buffalo's Airborne Networking and Communications Testbed (UB-ANC). To the best of our knowledge, UB-ANC is the first aerial networking platform that combines quadcopters capable of autonomous flight with sophisticated mission planning capabilities and flexible SDR-based transceivers, while also supporting off-the-shelf transceivers like Wi-Fi, Zigbee, and LTE. UB-ANC is designed to be modular and extensible in terms of both hardware and software, and it is built around popular open-source software and standards to facilitate its adoption. We have also developed the UB-ANC Emulator to enable UB-ANC's core software components to be debugged and tested in a software emulation before they are deployed in an actual UAV network. Although we present UB-ANC in the context of quadcopters, it can be used for other types of multirotors as well as helicopters, planes, boats, and rovers. UB-ANC and UB-ANC Emulator are open-source projects available via GitHub:\smallskip

{\tt ~~\url{https://github.com/jmodares/UB-ANC}}

{\tt ~~\url{https://github.com/jmodares/UB-ANC-Emulator}} \smallskip

With the UB-ANC hardware and software architectures designed and implemented, and three fully functional custom-built UB-ANC drones, we are now ready to do UAV networking experiments. We plan to investigate several problems leveraging UB-ANC's SDR transceivers and sophisticated mission planning capabilities including multi-UAV coverage path planning with connectivity constraints and positioning of aerial assets in air-to-air and air-to-ground ad hoc networks in order to optimize various network performance metrics, e.g., connectivity, throughput, and delay.

\vspace{-2pt}
\section{Acknowledgment}
ACKNOWLEDGEMENT OF SUPPORT AND DISCLAIMER: (a) The University at Buffalo acknowledges the U.S. Government's support in the publication of this paper. This material is based upon work funded by the US Air Force Research Laboratory under Grant No. FA8750-14-1-0073. (b) Any opinions, findings and conclusions or recommendations expressed in this material are those of the author(s) and do not necessarily reflect the views of AFRL. 

\vspace{-5pt}
\bibliographystyle{acm}
\bibliography{ref}
\balance

\end{document}